\documentclass[11pt]{article}

\usepackage{amsfonts,amssymb,chet,dsfont,graphicx,mathrsfs,tikz,cite}

\newcommand{\1}{\mathds{1}}

\newcommand{\ee}[3]{(\eta_{#1}\cdot\eta_{#2})^{#3}}
\newcommand{\e}[3]{\eta_{#1}^{#2_{#3}}}
\newcommand{\D}{\mathcal{D}}
\newcommand{\A}{\mathcal{A}}

\title{Conformal Differential Operator in Embedding\\Space and its Applications}

\author{Jean-Fran\c{c}ois Fortin$^{\ast,}$\email{jean-francois.fortin@phy.ulaval.ca} and Witold Skiba$^{\dagger,}$\email{witold.skiba@yale.edu}}

\affiliation{
$^\ast$D\'epartement de Physique, de G\'enie Physique et d'Optique\\Universit\'e Laval, Qu\'ebec, QC G1V 0A6, Canada\\
$^\dagger$Department of Physics, Yale University, New Haven, CT 06520, USA
}%Choices for affiliations $^{\ast,\dagger,\$,\S,\ddag,}$

\abstract{We develop techniques useful for obtaining conformal blocks in embedding space. We construct a unique differential operator in embedding space and use it to construct a function that will be an important ingredient in assembling conformal blocks. We show a number of relations that the components of conformal blocks satisfy and find invariance of our expressions under the dihedral group. }

\date{December 2016} %Uncomment this line for month to be fixed

\begin{document}

\maketitle

%\toc

%%%%%%%%%%%%%%%%%%%%%%%%%%%%%%%%%%%%%%%%%%%%%%%%%%
%%%%%%%%%%%%%%%%%%%%%%%%%%%%%%%%%%%%%%%%%%%%%%%%%%

\section{Introduction}\label{SecIntro}

Conformal field theories (CFTs) have many interesting properties and applications to various physical systems. The extension of Lorentz symmetry to conformal  symmetry provides quite stringent constraints on CFTs. The quantities of interest in CFTs are the spectrum of operators and their correlation functions. Symmetries restrict the form of two- and three-point correlation functions, such that the information about the dynamics is encoded in purely numerical coefficients. Because the operator product expansion (OPE) is convergent in CFTs, the OPE can be used to systematically reduce higher-point functions until only two- and three-point functions remain~\cite{Mack:1976pa}. 

Moreover, it was observed in the seventies that crossing symmetry of four-point functions can be used to constrain the possible values of operator dimensions and coefficients of three-point functions~\cite{Ferrara:1973yt,Polyakov:1974gs}. That observation proved to be very fruitful when applied to two-dimensional CFTs. When a four-point function is reduced by applying the OPE to pairs of points, symmetries reduce the problem to a sum over operators that are exchanged  in either the $s$, $t$, or $u$ channels. The expression for an individual operator exchange is governed by conformal symmetry and is called the conformal block. In larger number of dimensions, similar techniques have been used with much success only more recently~\cite{Rattazzi:2008pe}. In part, it took a lot longer to adopt bootstrap techniques beyond two dimensions because analytic information about conformal blocks has been rather limited with some of the first results obtained in \cite{Dolan:2000ut,Dolan:2003hv}. 

Conformal blocks are fully determined by symmetries. Yet, conformal symmetry acts non-linearly on the coordinates, which obscures consequences of the symmetry group. The most natural, linear, action of conformal symmetry in $d$ dimensions is on a $d+2$-dimensional space called the embedding space \cite{Dirac:1936fq}. While many results on CFTs were obtained using the embedding space, see for example \cite{Mack:1969rr,Weinberg:2010fx,Weinberg:2012mz}, the advantage of embedding space formulation has not been fully harnessed (articles in \cite{Ferrara:1971zy,Ferrara:1971vh,Ferrara:1972cq,Ferrara:1973eg} contain early important work on the OPE). 

In \cite{Fortin:2016lmf}, we outlined a program of deriving conformal blocks in the embedding space and showed how to obtain some already known results using our method. Here, we provide further ingredients that are necessary to implement conformal block calculations and the bootstrap program in the embedding space. There are two main parts of this article. In the next section, we investigate the differential operators that are consistent with the light-cone constraint of the embedding space and could appear on the right-hand side of the OPE. We show that all such operators can be generated from a single operator. We then investigate how this differential operator acts on the conformal cross ratios. In the following section, we show that the most general conformal block can be obtained from the action of the differential operator. We are concerned only with conformal blocks without any uncontracted Lorentz indices. Conformal blocks containing free Lorentz indices can be obtained by acting with more derivatives on the expression presented in this article. We also point out that the expression from which conformal blocks can be constructed is invariant under the dihedral group with 12 elements.

%%%%%%%%%%%%%%%%%%%%%%%%%%%%%%%%%%%%%%%%%%%%%%%%%%
%%%%%%%%%%%%%%%%%%%%%%%%%%%%%%%%%%%%%%%%%%%%%%%%%%

\section{Differential Operator}\label{SecDiff}

We denote the embedding space coordinate as $\eta^A$, where $\eta^A=(\eta^\mu,\eta^{d+1},\eta^{d+2})$. The embedding space is a projective space with $\eta^A$ and $\lambda \eta^A$ identified for $\lambda > 0$, and it is restricted to the light cone $\eta^2\equiv\eta^A \eta_A=0$.

The OPE expresses the product of two quasi-primary operators in terms of a series of quasi-primary operators and their descendants.  To generate the descendants, the OPE must thus include differential operators which act on the quasi-primary operators.  The only consistent differential operators with one derivative which are well defined on the light cone are
\eqn{\Theta=\eta^A\frac{\partial}{\partial\eta^A}\quad\quad\text{and}\quad\quad\mathcal{L}_{AB}=i\left(\eta_A\frac{\partial}{\partial\eta^B}-\eta_B\frac{\partial}{\partial\eta^A}\right).}[EqDiff1]
$\Theta$ is the homogeneity operator, while $\mathcal{L}_{AB}$ are the Lorentz generators in the $(d+2)$-dimensional embedding space, thus they satisfy the conformal algebra in $d$ dimensions.  With two derivatives, the only consistent differential operator which is not built from \eqref{EqDiff1} is the Thomas-Todorov operator~\cite{Dobrev:1975ru,Bailey:1994,Eastwood:2002su}
\eqn{\mathcal{K}_A=\left(\eta^B\frac{\partial}{\partial\eta^B}+\frac{d}{2}\right)\frac{\partial}{\partial\eta^A}-\frac{1}{2}\eta_A\frac{\partial}{\partial\eta_B}\frac{\partial}{\partial\eta^B}.}[EqDiff2]
Moreover, the non-vanishing commutation relations satisfied by the homogeneity operator, the conformal generators, and the Thomas-Todorov operator are
\eqn{
\begin{gathered}
{}[\mathcal{L}_{AB},\mathcal{L}_{CD}]=-i(g_{AC}\mathcal{L}_{BD}-g_{BC}\mathcal{L}_{AD}+g_{AD}\mathcal{L}_{CB}-g_{BD}\mathcal{L}_{CA}),\\
[\Theta,\mathcal{K}_A]=-\mathcal{K}_A,\qquad[\mathcal{L}_{AB},\mathcal{K}_C]=i(g_{BC}\mathcal{K}_A-g_{AC}\mathcal{K}_B).
\end{gathered}
}

Neither $\Theta$, $\mathcal{L}_{AB}$ nor $\mathcal{K}_A$ involve derivatives with respect to $\eta^2$ and are therefore well defined on the light cone. This can be seen by changing coordinates from $\eta^A=(\eta^\mu,\eta^{d+1},\eta^{d+2})$ to $\overline{\eta}^A=(x^\mu,k,\eta^2)$, where
\eqn{\eta^\mu=kx^\mu,\quad\quad\eta^{d+1}=\frac{\eta^2-k^2(1+x^2)}{2k},\quad\quad\eta^{d+2}=\frac{\eta^2+k^2(1-x^2)}{2k},}
or equivalently
\eqn{x^\mu=\frac{\eta^\mu}{-\eta^{d+1}+\eta^{d+2}},\quad\quad k=-\eta^{d+1}+\eta^{d+2},\quad\quad\eta^2=\eta^\mu\eta_\mu-(\eta^{d+1})^2+(\eta^{d+2})^2.}
In the new variables, one obtains
\eqna{
\frac{\partial}{\partial\eta^A}&=\frac{1}{k}[g_A^{\phantom{A}\mu}-(-g_A^{\phantom{A}d+1}+g_A^{\phantom{A}d+2})x^\mu]\frac{\partial}{\partial x^\mu}+(-g_A^{\phantom{A}d+1}+g_A^{\phantom{A}d+2})\frac{\partial}{\partial k}+2\eta_A\frac{\partial}{\partial\eta^2},\\
\eta_A\frac{\partial}{\partial\eta^B}&=\frac{\eta_A}{k}[g_B^{\phantom{B}\mu}-(-g_B^{\phantom{B}d+1}+g_B^{\phantom{B}d+2})x^\mu]\frac{\partial}{\partial x^\mu}+(-g_B^{\phantom{B}d+1}+g_B^{\phantom{B}d+2})\eta_A\frac{\partial}{\partial k}+2\eta_A\eta_B\frac{\partial}{\partial\eta^2},\\
\frac{\partial}{\partial\eta_B}\frac{\partial}{\partial\eta^B}&=\frac{1}{k^2}\frac{\partial}{\partial x_\mu}\frac{\partial}{\partial x^\mu}+4\left(k\frac{\partial}{\partial k}+\eta^2\frac{\partial}{\partial\eta^2}+\frac{d+2}{2}\right)\frac{\partial}{\partial\eta^2},
}
which can still be simplified on the light cone.  Hence
\eqna{
\Theta&=k\frac{\partial}{\partial k}+2\eta^2\frac{\partial}{\partial\eta^2},\\
\mathcal{L}_{AB}&=\frac{i}{k}\{\eta_A[g_B^{\phantom{B}\mu}-(-g_B^{\phantom{B}d+1}+g_B^{\phantom{B}d+2})x^\mu]-\eta_B[g_A^{\phantom{A}\mu}-(-g_A^{\phantom{A}d+1}+g_A^{\phantom{A}d+2})x^\mu]\}\frac{\partial}{\partial x^\mu}\\
&\phantom{=}\hspace{20pt}+i[\eta_A(-g_B^{\phantom{B}d+1}+g_B^{\phantom{B}d+2})-\eta_B(-g_A^{\phantom{A}d+1}+g_A^{\phantom{A}d+2})]\frac{\partial}{\partial k},\\
\mathcal{K}_A&=-\frac{\eta_A}{2k^2}\frac{\partial}{\partial x_\mu}\frac{\partial}{\partial x^\mu}+\frac{1}{k}[g_A^{\phantom{A}\mu}-(-g_A^{\phantom{A}d+1}+g_A^{\phantom{A}d+2})x^\mu]\left(k\frac{\partial}{\partial k}+2\eta^2\frac{\partial}{\partial\eta^2}+\frac{d}{2}\right)\frac{\partial}{\partial x^\mu}\\
&\phantom{=}\hspace{20pt}+(-g_A^{\phantom{A}d+1}+g_A^{\phantom{A}d+2})\left(k\frac{\partial}{\partial k}+2\eta^2\frac{\partial}{\partial\eta^2}+\frac{d+2}{2}\right)\frac{\partial}{\partial k}+2\eta_A\eta^2\frac{\partial}{\partial\eta^2}\frac{\partial}{\partial\eta^2},
}
which shows that $\Theta$, $\mathcal{L}_{AB}$ and $\mathcal{K}_A$ are well defined on the light cone.\footnote{Another way to ensure that a differential operator is well defined on the light cone is to check that its effect on $\eta^2=0$ is consistent.  For example, $\frac{\partial}{\partial\eta^A}\eta^2=2\eta_A\neq0$ thus $\frac{\partial}{\partial\eta^A}$ is not a well defined differential operator on the light cone.  On the other hand, $\Theta\eta^2=2\eta^2=0$ and $\mathcal{L}_{AB}\eta^2=2i(\eta_A\eta_B-\eta_B\eta_A)=0$ are consistent.}

Note that, due to the homogeneity condition of quasi-primary operators in embedding space, the differential operator $\Theta$ acts trivially.  Indeed, it does not generate descendants from the quasi-primary operators and can be discarded as a genuine differential operator.  Therefore, the only non-trivial differential operators which are well defined on the light cone must be made out of partial derivatives in the specific combinations $\mathcal{L}_{AB}$ or $\mathcal{K}_A$.

%%%%%%%%%%%%%%%%%%%%%%%%%%%%%%%%%%%%%%%%%%%%%%%%%%
%%%%%%%%%%%%%%%%%%%%%%%%%%%%%%%%%%%%%%%%%%%%%%%%%%

\subsection{Elementary Operator}

We now show that all non-trivial differential operators in the embedding space relevant for the OPE (\textit{i.e.} with two embedding space coordinates) can be obtained from a single operator.

Because the Casimir operators cannot generate descendant operators it is straightforward to conclude that all possible non-trivial differential operators can be written as combinations of $(\mathcal{L}^n)_{AB}$ for $n>0$ and $\mathcal{K}_A$ (since $4\mathcal{K}^A\mathcal{K}_A=\eta^2\partial^2\partial^2$ vanishes on the light cone).

It turns out that $(\mathcal{L}^n)_{AB}$ can be written in the following form $(\mathcal{L}^n)_{AB}=\eta_A\eta_B\partial^2S_n+\eta_A\partial_BT_n+\eta_B\partial_AU_n+g_{AB}V_n$, where the derivatives act  to the right and the operators $S_n$, $T_n$, $U_n$, and $V_n$ only depend on the homogeneity operator $\Theta$. The form of $(\mathcal{L}^n)_{AB}$ is preserved when $n$ is increased by one
\eqna{
(\mathcal{L}^{n+1})_{AB}&\equiv g^{CD} \mathcal{L}_{AC}(\mathcal{L}^n)_{DB}\\
&=\eta_A\eta_B\partial^2[i(d_E-2+\Theta)S_n+iU_n]+\eta_A\partial_B[i(d_E-2+\Theta)T_n+iU_n+iV_n]\\
&\phantom{=}\hspace{20pt}+\eta_B\partial_A[i(1-\Theta)U_n-iV_n]+g_{AB}(-i\Theta U_n),
}
which gives the following relations
\eqna{
S_{n+1}&=i(d_E-2+\Theta)S_n+iU_n,\\
T_{n+1}&=i(d_E-2+\Theta)T_n+iU_n+iV_n,\\
U_{n+1}&=i(1-\Theta)U_n-iV_n,\\
V_{n+1}&=-i\Theta U_n,
}
where $d_E=d+2$ is the dimension of the embedding space. One can then show that the differential operators $(\mathcal{L}^n)_{AB}=(\mathcal{L}^2)_{AB}A_n+\mathcal{L}_{AB}B_n+g_{AB}C_n$, where $A_n$, $B_n$, and $C_n$ are again functions of the homogeneity operator $\Theta$.  Indeed, for $n=1$ one obtains $\mathcal{L}_{AB}$ by setting $S_1=0$, $T_1=i$, $U_1=-i$, and $V_1=0$. Given these initial values the recursion relations are satisfied by 
\eqna{
S_n&=\frac{-i^n[(1+\Theta)(d_E-2+\Theta)^n+(d_E-3+\Theta)(-\Theta)^n-(d_E-2+2\Theta)]}{(1+\Theta)(d_E-3+\Theta)(d_E-2+2\Theta)},\\
T_n&=\frac{i^n[(1+\Theta)(d_E-4+2\Theta)(d_E-2+\Theta)^n-2(d_E-3+\Theta)(-\Theta)^n+(1-\Theta)(d_E-2+2\Theta)]}{(1+\Theta)(d_E-3+\Theta)(d_E-2+2\Theta)},\\
U_n&=\frac{-i^n[1-(-\Theta)^n]}{1+\Theta},\\
V_n&=\frac{i^n\Theta[1-(-\Theta)^{n-1}]}{1+\Theta}.
}
Therefore, $(\mathcal{L}^n)_{AB}$ is simply given by
\eqna{
(\mathcal{L}^n)_{AB}&=\eta_A\eta_B\partial^2S_n+\eta_A\partial_BT_n+\eta_B\partial_AU_n+g_{AB}V_n\\
&=\left[\eta_A\eta_B\partial^2+\eta_A\partial_B\frac{T_n+U_n-(1-\Theta)S_n}{S_n}+\eta_B\partial_A(1-\Theta)+g_{AB}(-\Theta)\right]S_n\\
&\phantom{=}\hspace{20pt}-(\eta_A\partial_B-\eta_B\partial_A)[U_n-(1-\Theta)S_n]+g_{AB}(V_n+\Theta S_n)\\
&=[\eta_A\eta_B\partial^2-\eta_A\partial_B(d_E-3+\Theta)+\eta_B\partial_A(1-\Theta)+g_{AB}(-\Theta)]S_n\\
&\phantom{=}\hspace{20pt}+i(\eta_A\partial_B-\eta_B\partial_A)[iU_n-i(1-\Theta)S_n]+g_{AB}(V_n+\Theta S_n)\\
&=(\mathcal{L}^2)_{AB}A_n+\mathcal{L}_{AB}B_n+g_{AB}C_n,
}
where $A_n=S_n$, $B_n=iU_n-i(1-\Theta)S_n$, and $C_n=V_n+\Theta S_n$.  Since the homogeneity operator commutes with the conformal generator, all non-trivial differential operators can be expressed in terms of linear combinations of $\mathcal{L}_{AB}$, $(\mathcal{L}^2)_{AB}$ and $\mathcal{K}_A$.  Indeed, since
\eqna{
\mathcal{L}_{AB}\mathcal{K}^B&=-i\Theta\mathcal{K}_A+\frac{i}{2}\eta^2\partial_A\partial^2,\\
(\mathcal{L}^2)_{AB}\mathcal{K}^B&=-\Theta^2\mathcal{K}_A+\eta^2\left(\mathcal{K}_A-\frac{d-2}{2}\partial_A\right)\partial^2,
}
on the light cone $\eta^2=0$, all remaining products are linear combinations of $\mathcal{L}_{AB}$, $(\mathcal{L}^2)_{AB}$ and $\mathcal{K}_A$ with different factors which are functions of $\Theta$.  An analog result has been obtained in \cite{Eastwood:2002su} by studying the symmetries of the Laplacian operator. 

The OPE involves operators at two points on the light cone, $\eta_1$ and $\eta_2$.  We use the convention that the operators on the right-hand side of the OPE are located at $\eta_2$, therefore the differential operators present in the OPE must act at $\eta_2$.  There are then three differential operators without contractions given by
\eqna{
\mathcal{L}_{2AB}&=i(\eta_{2A}\partial_{2B}-\eta_{2B}\partial_{2A}),\\
(\mathcal{L}_2^2)_{AB}&=\eta_{2A}\eta_{2B}\partial_2^2-\eta_{2A}\partial_{2B}(d_E-3+\Theta_2)+\eta_{2B}\partial_{2A}(1-\Theta_2)-g_{AB}\Theta_2,\\
\mathcal{K}_{2A}&=\left(\Theta_2+\frac{d}{2}\right)\partial_{2A}-\frac{1}{2}\eta_{2A}\partial_2^2.
}
There are also five simple differential operators which can be constructed with one contraction from the conformal generators,
\eqna{
(\eta_1\cdot\mathcal{L}_2)^A&=i\ee{1}{2}{}\partial_2^A-i\e{2}{A}{}\eta_1\cdot\partial_2,\\
(\eta_1\cdot\mathcal{L}_2^2)^A&=\ee{1}{2}{}\e{2}{A}{}\partial_2^2-\ee{1}{2}{}\partial_2^A(d_E-3+\Theta_2)+\e{2}{A}{}\eta_1\cdot\partial_2(1-\Theta_2)-\e{1}{A}{}\Theta_2,\\
(\eta_2\cdot\mathcal{L}_2^n)^A&=\e{2}{A}{}(-i\Theta_2)^n,\\
\eta_1\cdot\mathcal{K}_2&=\left(\Theta_2+\frac{d}{2}\right)\eta_1\cdot\partial_2-\frac{1}{2}\ee{1}{2}{}\partial_2^2,\\
\eta_2\cdot\mathcal{K}_2&=\Theta_2\left(\Theta_2-1+\frac{d}{2}\right).
}
The third and last operators can be disregarded since they depend on the homogeneity operator only.  Due to the antisymmetry of the conformal generators, there are only two differential operators with two contractions which are given by
\eqna{
\eta_1\cdot\mathcal{L}_2^2\cdot\eta_1&\equiv\e{1}{A}{}\e{1}{B}{}(\mathcal{L}_2^2)_{AB}=\ee{1}{2}{2}\partial_2^2-\ee{1}{2}{}\eta_1\cdot\partial_2(d_E-4+2\Theta_2),\\
\eta_1\cdot\mathcal{L}_2^2\cdot\eta_2&\equiv\e{1}{A}{}\e{2}{B}{}(\mathcal{L}_2^2)_{AB}=\e{1}{A}{}\e{2}{B}{} \left[ (\mathcal{L}_2^2)_{BA} - i (d_E-2) (\mathcal{L}_2)_{BA}\right]\\
&=-\ee{1}{2}{} \left[ \Theta_2^2 + (d_E-2) \Theta_2 \right].
}
Once again, the second operator is trivial because it is expressed in terms of the homogeneity operator only.  

There remain seven operators that could be useful: three with no contractions, three with one contraction, and one with two contractions.  Of those seven, only one is independent since
\eqna{
\mathcal{L}_{2AB}&=\frac{1}{\ee{1}{2}{}}[\eta_{2A}(\eta_1\cdot\mathcal{L}_2)_B-\eta_{2B}(\eta_1\cdot\mathcal{L}_2)_A],\\
(\mathcal{L}_2^2)_{AB}&=\mathcal{L}_{2AC}\mathcal{L}_{2\phantom{C}B}^{\phantom{2}C},\\
\mathcal{K}_{2A}&=-\frac{1}{2\ee{1}{2}{}}[(\eta_1\cdot\mathcal{L}_2^2)_A+i(\eta_1\cdot\mathcal{L}_2)_A(\Theta_2-1)+\eta_{1A}\Theta_2],\\
(\eta_1\cdot\mathcal{L}_2^2)^A&=\frac{\e{2}{A}{}}{\ee{1}{2}{}}\eta_1\cdot\mathcal{L}_2^2\cdot\eta_1+i(\eta_1\cdot\mathcal{L}_2)^A(d_E-3+\Theta_2)-\e{1}{A}{}\Theta_2,\\
\eta_1\cdot\mathcal{L}_2^2\cdot\eta_1&=-(\eta_1\cdot\mathcal{L}_2)^A(\eta_1\cdot\mathcal{L}_2)_A.
}
Therefore, there is only one non-trivial differential operator which plays a significant role in the OPE, and it is chosen to be
\eqna{
\D_{12}^A&\equiv\frac{1}{\ee{1}{2}{\frac{1}{2}}}[-i(\eta_1\cdot\mathcal{L}_2)^A-\e{1}{A}{}\Theta_2]=\frac{1}{\ee{1}{2}{\frac{1}{2}}}[\ee{1}{2}{}\partial_2^A-\e{2}{A}{}\eta_1\cdot\partial_2-\e{1}{A}{}\eta_2\cdot\partial_2]\\
&=\ee{1}{2}{\frac{1}{2}}\A_{12}^{AB}\partial_{2B},\quad\quad\text{where}\quad\quad\A_{12}^{AB}=\frac{1}{\ee{1}{2}{}}[\ee{1}{2}{}g^{AB}-\e{1}{A}{}\e{2}{B}{}-\e{1}{B}{}\e{2}{A}{}].
}[EqDO]
The prefactor $\ee{1}{2}{\frac{1}{2}}$ is introduced for future convenience.  All remaining differential operators built from \eqref{EqDiff1} and \eqref{EqDiff2} can be expressed in terms of the differential operator \EqDO, thanks to the extra embedding space coordinate $\eta_1$.

Because $\A_{12}^{AB}$ is transverse in each of its indices, the differential operator in \EqDO has a special status and satisfies several interesting identities.  The most important ones, which can all be proven by induction, are given below. First, note that $\A_{12}^{AB}=\A_{12}^{BA}=\A_{21}^{AB}=\A_{21}^{BA}$ and
\eqn{\eta_{1A}\A_{12}^{AB}=\eta_{2A}\A_{12}^{AB}=0,\quad\quad\A_{12}^{AC}\A_{12C}^{\phantom{12C}B}=\A_{12}^{AB}.}
Therefore, $\D_{12}^A$ trivially satisfies
\eqn{\e{1}{A}{}\D_{12A}=\e{2}{A}{}\D_{12A}=0\quad\quad\text{and}\quad\quad\A_{12}^{AB}\D_{12B}=\D_{12}^A,}
which imply that
\eqn{\D_{12}^A\ee{1}{2}{\alpha}-\ee{1}{2}{\alpha}\D_{12}^A=0.}[EqDAeta]
The non-trivial commutation relations between $\D_{12}^A$ and $\Theta_{1,2}$ are given by
\eqn{
\begin{array}{c}
[\D_{12}^A,\D_{12}^B]=\frac{1}{\ee{1}{2}{\frac{1}{2}}}(\e{1}{A}{}\D_{12}^B-\e{1}{B}{}\D_{12}^A),\quad\quad[\Theta_1,\D_{12}^A]=\frac{1}{2}\D_{12}^A,\quad\quad[\Theta_2,\D_{12}^A]=-\frac{1}{2}\D_{12}^A,\vspace{8pt}\\
\e{2}{A}{}\D_{12}^B-\D_{12}^B\e{2}{A}{}=-\ee{1}{2}{\frac{1}{2}}\A_{12}^{AB}.
\end{array}
}[EqDA]
The second and third equations of \EqDA imply that $\D_{12}^A$ has well defined degrees of homogeneity in $\eta_1$ and $\eta_2$.  

Squaring $\D_{12}^A$ gives the scalar differential operator 
\vspace{-10pt}
\small
\eqn{\D_{12}^2\equiv\D_{12}^A\D_{12A}=\frac{1}{\ee{1}{2}{}}\eta_1\cdot\mathcal{L}_2^2\cdot\eta_1=\ee{1}{2}{}\partial_2^2-\eta_1\cdot\partial_2(d_E-4+2\Theta_2)=\ee{1}{2}{}\partial_2^2-(d_E-2+2\Theta_2)\eta_1\cdot\partial_2,}
\normalsize
for which
\eqn{
\begin{array}{c}
[\D_{12}^A,\D_{12}^2]=\frac{2}{\ee{1}{2}{\frac{1}{2}}}\e{1}{A}{}\D_{12}^2,\quad\quad[\Theta_1,\D_{12}^2]=\D_{12}^2,\quad\quad[\Theta_2,\D_{12}^2]=-\D_{12}^2,\vspace{8pt}\\
\e{2}{A}{}\D_{12}^2-\D_{12}^2\e{2}{A}{}=-2\ee{1}{2}{\frac{1}{2}}\D_{12}^A+(d_E-2)\e{1}{A}{}.
\end{array}
}[EqDAsq]
With the help of fractional calculus, the scalar differential operator can naturally be applied $h/2$ times to a quasi-primary operator, where $h\in\mathbb{R}$.  This  generalization will be needed because conformal dimensions are real numbers.  Commutation relations in \EqDAsq can be generalized to arbitrary $h$ as follows
\eqn{
\begin{array}{c}
[\D_{12}^A,\D_{12}^h]=\frac{h}{\ee{1}{2}{\frac{1}{2}}}\e{1}{A}{}\D_{12}^h,\quad\quad[\Theta_1,\D_{12}^h]=\frac{h}{2}\D_{12}^h,\quad\quad[\Theta_2,\D_{12}^h]=-\frac{h}{2}\D_{12}^h,\vspace{8pt}\\
\e{2}{A}{}\D_{12}^h-\D_{12}^h\e{2}{A}{}=-h\ee{1}{2}{\frac{1}{2}}\D_{12}^A\D_{12}^{h-2}+\frac{h}{2}(d_E+h-4)\e{1}{A}{}\D_{12}^{h-2}.
\end{array}
}[EqDAsqh]
Finally, the commutation relations for the conformal generators with the embedding space coordinates and the differential operators are
\eqn{
\begin{array}{c}
\mathcal{L}_{AB}\e{}{C}{}-\e{}{C}{}\mathcal{L}_{AB}=-(S_{AB})_{\phantom{C}D}^C\e{}{D}{},\vspace{8pt}\\
{[}\mathcal{L}_{1AB}+\mathcal{L}_{2AB},\D_{12}^C]=-(S_{AB})_{\phantom{C}D}^C\D_{12}^D,\vspace{8pt}\\
{[}\mathcal{L}_{1AB}+\mathcal{L}_{2AB},\D_{12}^h]=0,
\end{array}
}[EqC]
where $(S_{AB})_{\phantom{C}D}^C$ are the $SO(d,2)$ generators in the vector representation.

%%%%%%%%%%%%%%%%%%%%%%%%%%%%%%%%%%%%%%%%%%%%%%%%%%
%%%%%%%%%%%%%%%%%%%%%%%%%%%%%%%%%%%%%%%%%%%%%%%%%%

\subsection{Change of Variables}

It will obviously be useful to express the operator $\D_{12}^A$ in terms of the conformal ratios
\eqn{u=\frac{\ee{1}{2}{}\ee{3}{4}{}}{\ee{1}{3}{}\ee{2}{4}{}}\quad\quad\text{and}\quad\quad v=\frac{\ee{1}{4}{}\ee{2}{3}{}}{\ee{1}{3}{}\ee{2}{4}{}}.}[EqCR]
Since $\D_{12}^A$ commutes with $\ee{1}{2}{\alpha}$, as shown in \EqDAeta, the differential operator acts non-trivially only on $\ee{2}{3}{}$ and $\ee{2}{4}{}$ and one obtains 
\eqna{
\D_{12}^Au^\alpha-u^\alpha\D_{12}^A&=\ee{1}{2}{\frac{1}{2}}\A_{12}^{AB}(\partial_{2B}u^\alpha)=-\alpha u^\alpha\ee{1}{2}{\frac{1}{2}}\A_{12}^{AB}\frac{\eta_{4B}}{\ee{2}{4}{}},\\
\D_{12}^Av^\beta-v^\beta\D_{12}^A&=\ee{1}{2}{\frac{1}{2}}\A_{12}^{AB}(\partial_{2B}v^\beta)=\beta v^\beta\ee{1}{2}{\frac{1}{2}}\A_{12}^{AB}\left[\frac{\eta_{3B}}{\ee{2}{3}{}}-\frac{\eta_{4B}}{\ee{2}{4}{}}\right].
}
The conformal ratios have vanishing degrees of homogeneity, so it is of interest to redefine the differential operator $\D_{12}^A$ such that it is homogenous with respect to all four coordinates.   The rescaled  operator is defined as
\eqn{\D_{(u,v)}^A=\frac{\ee{1}{2}{\frac{1}{2}}\ee{3}{4}{\frac{1}{2}}}{\ee{1}{3}{\frac{1}{2}}\ee{1}{4}{\frac{1}{2}}}\D_{12}^A\quad\quad\text{and}\quad\quad\D_{(u,v)}=\D^A\D_A=\frac{\ee{1}{2}{}\ee{3}{4}{}}{\ee{1}{3}{}\ee{1}{4}{}}\D_{12}^2.}
The scalar operator $\D_{(u,v)}$ is more suited to act on functions of the conformal ratios since its action results in other functions of the conformal ratios only.  When $\D_{(u,v)}^A$ and $\D_{(u,v)}$ act on functions of conformal ratios only, one can change variables such that the derivatives are with respect to $u$ and $v$ only
\eqna{
\D_{(u,v)}^A&=\left(\D_{(u,v)}^Au\right)\partial_u+\left(\D_{(u,v)}^Av\right)\partial_v,\\
\D_{(u,v)}&=\D_{(u,v)}^A\left[\left(\D_{(u,v)A}u\right)\partial_u+\left(\D_{(u,v)A}v\right)\partial_v\right]\\
&=\left(\D_{(u,v)}^Au\right)\left(\D_{(u,v)A}u\right)\partial_u^2+2\left(\D_{(u,v)}^Au\right)\left(\D_{(u,v)A}v\right)\partial_u\partial_v\\
&\phantom{=}\hspace{20pt}+\left(\D_{(u,v)}^Av\right)\left(\D_{(u,v)A}v\right)\partial_v^2+\left(\D_{(u,v)}^2u\right)\partial_u+\left(\D_{(u,v)}^2v\right)\partial_v.
}
It is then straightforward to verify that
\eqna{
\D_{(u,v)}^A&=-\frac{\ee{1}{2}{}\ee{3}{4}{\frac{1}{2}}}{\ee{1}{3}{\frac{1}{2}}\ee{1}{4}{\frac{1}{2}}}\A_{12}^{AB}\left\{\frac{\eta_{4B}}{\ee{2}{4}{}}u\partial_u-\left[\frac{\eta_{3B}}{\ee{2}{3}{}}-\frac{\eta_{4B}}{\ee{2}{4}{}}\right]v\partial_v\right\},\\
\D_{(u,v)}&=(-2)\left\{u^3\partial_u^2+u^2(u+v-1)\partial_u\partial_v+u^2v\partial_v^2-\left(\tfrac{d}{2}-2\right)u^2\partial_u+u\left[u+\left(\tfrac{d}{2}-1\right)(1-v)\right]\partial_{v}\right\}.
}[EqD]

Because of transversality of $\A_{12}^{AB}$, $\D_{(u,v)}^A$ can only be contracted with either $\eta_{3A}$ or $\eta_{4A}$. These contractions give the following new operators, where their respective pre-factors were chosen for homogeneity, 
\eqna{
\D_{(u)}&=-\frac{\ee{1}{3}{\frac{1}{2}}\eta_{4A}}{\ee{1}{4}{\frac{1}{2}}\ee{3}{4}{\frac{1}{2}}}\D_{(u,v)}^A=-2u\partial_u-(u+v-1)\partial_v,\\
\D_{(v)}&=-\frac{\ee{1}{4}{\frac{1}{2}}\eta_{3A}}{\ee{1}{3}{\frac{1}{2}}\ee{3}{4}{\frac{1}{2}}}\D_{(u,v)}^A=u(u-v-1)\partial_u+v(u-v+1)\partial_v.
}
Their algebra is 
\eqna{
[\D_{(u)},\D_{(v)}]&=\D_{(u)}-\D_{(v)},\\
[\D_{(u)},\D_{(u,v)}^h]&=-2h\D_{(u,v)}^h,\\
[\D_{(v)},\D_{(u,v)}^h]&=-2h\D_{(u,v)}^h,
}
and it can be obtained from \eqref{EqDA}, \eqref{EqDAsq} and \eqref{EqDAsqh}.  Moreover, they satisfy several important properties, for example
\eqna{
\D_{(u)}u^\alpha-u^\alpha\D_{(u)}&=(-2\alpha)u^\alpha,\\
\D_{(v)}\left(\frac{u}{v}\right)^\beta-\left(\frac{u}{v}\right)^\beta\D_{(v)}&=(-2\beta)\left(\frac{u}{v}\right)^\beta,\\
\D_{(u,v)}u^\alpha-u^\alpha\D_{(u,v)}&=(-2\alpha)u^{\alpha+1}\left(\alpha+1-\tfrac{d}{2}-\D_{(u)}\right),\\
\D_{(u,v)}\left(\frac{u}{v}\right)^\beta-\left(\frac{u}{v}\right)^\beta\D_{(u,v)}&=(-2\beta)\left(\frac{u}{v}\right)^{\beta+1}\left(\beta+1-\tfrac{d}{2}-\D_{(v)}\right).
}
A further generalization of these properties can be obtained using the general Leibniz rule and recursion  
\eqna{
\D_{(u)}^hu^\alpha&=\sum_{i\geq0}\genfrac{(}{)}{0pt}{}{h}{i}(-2\alpha)^iu^\alpha\D_{(u)}^{h-i},\\
\D_{(v)}^h\left(\frac{u}{v}\right)^\beta&=\sum_{i\geq0}\genfrac{(}{)}{0pt}{}{h}{i}(-2\beta)^i\left(\frac{u}{v}\right)^\beta\D_{(v)}^{h-i},\\
\D_{(u,v)}^hu^\alpha&=\sum_{i\geq0}\genfrac{(}{)}{0pt}{}{h}{i}(-2)^i(\alpha)_iu^{\alpha+i}\left(\alpha-h+i+1-\tfrac{d}{2}-\D_{(u)}\right)_i\D_{(u,v)}^{h-i},\\
\D_{(u,v)}^h\left(\frac{u}{v}\right)^\beta&=\sum_{i\geq0}\genfrac{(}{)}{0pt}{}{h}{i}(-2)^i(\beta)_i\left(\frac{u}{v}\right)^{\beta+i}\left(\beta-h+i+1-\tfrac{d}{2}-\D_{(v)}\right)_i\D_{(u,v)}^{h-i},
}
where $(\ldots)_i$ denotes the Pochhammer symbol. With a slight abuse of notation, the arguments of the Pochhammer symbols contain not just pure numbers, but sometimes operators as well. 
The terms with no derivatives on the right-hand sides of the equations above give the action of  the derivative operators on the powers of $u$ and $u/v$ alone
\eqna{
\D_{(u)}^hu^\alpha&=(-2\alpha)^hu^\alpha,\\
\D_{(v)}^h\left(\frac{u}{v}\right)^\beta&=(-2\beta)^h\left(\frac{u}{v}\right)^\beta,\\
\D_{(u,v)}^hu^\alpha&=(-2)^h(\alpha)_h(\alpha+1-d/2)_hu^{\alpha+h},\\
\D_{(u,v)}^h\left(\frac{u}{v}\right)^\beta&=(-2)^h(\beta)_h(\beta+1-d/2)_h\left(\frac{u}{v}\right)^{\beta+h},
}[EqDGF]
which we quote for future reference.

%%%%%%%%%%%%%%%%%%%%%%%%%%%%%%%%%%%%%%%%%%%%%%%%%%
%%%%%%%%%%%%%%%%%%%%%%%%%%%%%%%%%%%%%%%%%%%%%%%%%%

\section{Master Function}\label{SecMF}

We now turn to constructing the most general form of a conformal block  and show how to apply the derivatives developed in the previous section. We assume that the OPE is used once at points $1$ and $2$ and another time at points $3$ and $4$. The most general conformal block will depend on the Lorentz representations of both the external and exchange operators. The OPE will contain two types of derivatives: scalar derivatives raised to real powers, that correspond to scaling dimensions, and vector derivatives with integer powers that account for the operator spin. In this section we are concerned with the most general action of the scalar derivatives. Obtaining general conformal blocks will require proper accounting for the spins of the operators, but the answer can be expressed in terms of the function $F_d$ described below and derivatives acting on it. The details will be presented in a future publication.  

The numbers of embedding space positions and derivatives $\D_{12}^A$ and $\D_{34}^A$ that are not contracted are finite integer numbers that depend on the Lorentz representations of external operators. Such positions and derivatives with free Lorentz indices can be commuted to the left of the derivatives $\D_{12}^2$ and $\D_{34}^2$ with the help of the commutation relations \eqref{EqDAsqh}. Therefore, the most general contribution to any conformal block without free embedding space vector indices is
\eqna{
F_d^{(p,q;r;s,t)}\ee{i}{j}{}&=\D_{12}^{2p}\D_{34}^{2q}\ee{2}{4}{-r}\ee{2}{3}{s}\ee{1}{4}{t}\\
&=\frac{\ee{1}{3}{r+p+q}\ee{1}{4}{t-q-s+p}}{\ee{1}{2}{r-s+p}\ee{3}{4}{r-s+p}}\D_{(u,v)}^p\left(\frac{u}{v}\right)^{t-s-q}\D_{(u,v)}^q\left(\frac{u}{v}\right)^{r-t}v^r.
}[EqFpq]
From its definition and the fact that the derivatives $\D_{12}^2$ and $\D_{34}^2$ commute, it is clear that $F_d^{(p,q;r;s,t)}\ee{i}{j}{}$ can also be expressed as
\eqna{
F_d^{(p,q;r;s,t)}\ee{i}{j}{}&=\D_{12}^{2p}\D_{34}^{2q}\ee{2}{4}{-r}\ee{2}{3}{s}\ee{1}{4}{t}\\
&=\frac{\ee{1}{3}{r+p+q}\ee{2}{3}{s-p-t+q}}{\ee{1}{2}{r-t+q}\ee{3}{4}{r-t+q}}\D_{(u,v)}^q\left(\frac{u}{v}\right)^{s-t-p}\D_{(u,v)}^p\left(\frac{u}{v}\right)^{r-s}v^r.
}[EqFqp]

The two forms for $F_d^{(p,q;r;s,t)}\ee{i}{j}{}$ \eqref{EqFpq} and \eqref{EqFqp} and the behavior \eqref{EqDGF} of the derivative $\D_{(u,v)}$ as $u\to0$ and $v\to1$ suggest writing $F_d^{(p,q;r;s,t)}\ee{i}{j}{}$ as
\eqna{
F_d^{(p,q;r;s,t)}\ee{i}{j}{}&=(-2)^{p+q}(r-s)_p(r-t)_q(r-s+1-d/2)_p(r-t+1-d/2)_q\\
&\phantom{=}\hspace{20pt}\times\frac{\ee{1}{3}{r+p+q}\ee{1}{4}{t-q-s+p}}{\ee{1}{2}{r-s+p}\ee{3}{4}{r-s+p}}\left(\frac{u}{v}\right)^{r-s+p}H_d^{(p,q;r;s,t)}(u,v)\\
&=(-2)^{p+q}(r-s)_p(r-t)_q(r-s+1-d/2)_p(r-t+1-d/2)_q\\
&\phantom{=}\hspace{20pt}\times\frac{\ee{1}{3}{r+p+q}\ee{2}{3}{s-p-t+q}}{\ee{1}{2}{r-t+q}\ee{3}{4}{r-t+q}}\left(\frac{u}{v}\right)^{r-t+q}H_d^{(q,p;r;t,s)}(u,v),
}
where the function $H_d^{(p,q;r;s,t)}(u,v)$ is given by
\eqn{H_d^{(p,q;r;s,t)}(u,v)=\frac{\left(\frac{u}{v}\right)^{-(r-s+p)}\D_{(u,v)}^p\left(\frac{u}{v}\right)^{t-s-q}\D_{(u,v)}^q\left(\frac{u}{v}\right)^{r-t}v^r}{(-2)^{p+q}(r-s)_p(r-t)_q(r-s+1-d/2)_p(r-t+1-d/2)_q}.}[EqHGF]
The power of $\frac{u}{v}$ and the normalization are chosen such that $H_d^{(p,q;r;s,t)}(0,1)=1$.  Note that the equivalence of the two forms \eqref{EqFpq} and \eqref{EqFqp} for the function $F_d^{(p,q;r;s,t)}\ee{i}{j}{}$ implies the symmetry $H_d^{(p,q;r;s,t)}(u,v)=H_d^{(q,p;r;t,s)}(u,v)$.

To proceed, it is convenient to introduce another function,
\eqn{G_d^{(q;r;t)}(u,v)=\frac{\left(\frac{u}{v}\right)^{-(r-t+q)}\D_{(u,v)}^q\left(\frac{u}{v}\right)^{r-t}v^r}{(-2)^q(r-t)_q(r-t+1-d/2)_q},}[EqGGF]
in addition to function $H_d^{(p,q;r;s,t)}(u,v)$.  Again, from the behavior of $\D_{(u,v)}$ as $u\to0$ and $v\to1$ one concludes that $G_d^{(q;r;t)}(0,1)=1$.  Having defined $G_d^{(q;r;t)}(u,v)$, we can re-express $H_d^{(p,q;r;s,t)}(u,v)$ as the derivative operator acting on $G_d^{(q;r;t)}(u,v)$
\eqn{H_d^{(p,q;r;s,t)}(u,v)=\frac{\left(\frac{u}{v}\right)^{-(r-s+p)}\D_{(u,v)}^p\left(\frac{u}{v}\right)^{r-s}G_d^{(q;r;t)}(u,v)}{(-2)^p(r-s)_p(r-s+1-d/2)_p}.}[EqHinG]

It turns out that $G_d^{(q;r;t)}(u,v)$ and $H_d^{(p,q;r;s,t)}(u,v)$ have several simple limits. These are obtained using  \eqref{EqDGF} and are valid for arbitrary values of the remaining parameters. 
First, $G_d^{(q;r;t)}(u,v)$ obeys
\eqna{
G_d^{(0;r;t)}(u,v)&=v^r,\\
G_d^{(q;0;t)}(u,v)&=1,\\
G_d^{(q;r;0)}(u,v)&=v^{r+q},
}
while $H_d^{(p,q;r;s,t)}(u,v)$ satisfies 
\eqna{
H_d^{(0,q;r;s,t)}(u,v)&=G_d^{(q;r;t)}(u,v),\\
H_d^{(p,0;r;s,t)}(u,v)&=\frac{\left(\frac{u}{v}\right)^{-(r-s+p)}\D_{(u,v)}^p\left(\frac{u}{v}\right)^{r-s}v^r}{(-2)^p(r-s)_p(r-s+1-d/2)_p}=G_d^{(p;r;s)}(u,v),\\
H_d^{(p,q;0;s,t)}(u,v)&=\frac{\left(\frac{u}{v}\right)^{-(-s+p)}\D_{(u,v)}^p\left(\frac{u}{v}\right)^{-s}}{(-2)^p(-s)_p(-s+1-d/2)_p}=1,\\
H_d^{(p,q;r;0,t)}(u,v)&=\frac{\left(\frac{u}{v}\right)^{-(r+p)}\D_{(u,v)}^p\left(\frac{u}{v}\right)^rG_d^{(q;r;t)}(u,v)}{(-2)^p(r)_p(r+1-d/2)_p},\\
H_d^{(p,q;r;s,0)}(u,v)&=\frac{\left(\frac{u}{v}\right)^{-(r-s+p)}\D_{(u,v)}^p\left(\frac{u}{v}\right)^{r-s}v^{r+q}}{(-2)^p(r-s)_p(r-s+1-d/2)_p}=G_d^{(p;r+q;s+q)}(u,v).
}
The symmetry $H_d^{(p,q;r;s,t)}(u,v)=H_d^{(q,p;r;t,s)}(u,v)$ then implies the following identity for $G_d^{(q;r;t)}(u,v)$
\eqn{G_d^{(p;r+q;s+q)}(u,v)=\frac{\left(\frac{u}{v}\right)^{-(r+q)}\D_{(u,v)}^q\left(\frac{u}{v}\right)^rG_d^{(p;r;s)}(u,v)}{(-2)^q(r)_q(r+1-d/2)_q\\}.}
Setting $s=0$, the identity above leads to $G_d^{(p;r+q;q)}(u,v)=G_d^{(q;r+p;p)}(u,v)$. Renaming the parameters gives $G_d^{(q;r;t)}(u,v)=G_d^{(t;r-t+q;q)}(u,v)$.

Several other special cases can be obtained from the definition \eqref{EqHGF} and the symmetry.  For example, one has
\eqn{
H_d^{(s-t,q;r;s,t)}(u,v)=G_d^{(q+s-t;r;s)}(u,v).}
Moreover, the relation $G_d^{(q;r;t)}(u,v)=G_d^{(t;r-t+q,q)}(u,v)$ implies for $H_d^{(p,q;r;s,t)}$
\eqna{
H_d^{(p,q;r;s,t)}(u,v)&=\frac{\left(\frac{u}{v}\right)^{-(r-s+p)}\D_{(u,v)}^p\left(\frac{u}{v}\right)^{r-s}G_d^{(q;r;t)}(u,v)}{(-2)^p(r-s)_p(r-s+1-d/2)_p}\\
&=\frac{\left(\frac{u}{v}\right)^{-(r-s+p)}\D_{(u,v)}^p\left(\frac{u}{v}\right)^{r-s}G_d^{(t;r-t+q;q)}(u,v)}{(-2)^p(r-s)_p(r-s+1-d/2)_p}\\
&=\frac{\left(\frac{u}{v}\right)^{-(r-s+p)}\D_{(u,v)}^p\left(\frac{u}{v}\right)^{r-t+q-(s-t+q)}G_d^{(t;r-t+q;q)}(u,v)}{(-2)^p(r-s)_p(r-s+1-d/2)_p}\\
&=H_d^{(p,t;r-t+q;s-t+q,q)}(u,v).
}
This transformation is consistent with the special cases already mentioned and allows other relations to be found.

From their definitions, \eqref{EqGGF} and \eqref{EqHGF}, $G_d^{(q;r;t)}(u,v)$ and $H_d^{(p,q;r;s,t)}(u,v)$ satisfy the following recurrence relations
\eqna{
G_d^{(q+k;r;t)}(u,v)&=\frac{\left(\frac{u}{v}\right)^{-(r-t+q+k)}\D_{(u,v)}^k\left(\frac{u}{v}\right)^{r-t+q}G_d^{(q;r;t)}(u,v)}{(-2)^k(r-t+q)_k(r-t+q+1-d/2)_k},\\
H_d^{(p+k,q;r;s,t)}(u,v)&=\frac{\left(\frac{u}{v}\right)^{-(r-s+p+k)}\D_{(u,v)}^k\left(\frac{u}{v}\right)^{r-s+p}H_d^{(p,q;r;s,t)}(u,v)}{(-2)^k(r-s+p)_k(r-s+p+1-d/2)_k}.
}[EqRR]
The recurrence relations \eqref{EqRR} are exactly the same, but the two functions are different because of the initial conditions. Indeed, $G_d^{(0;r;t)}(u,v)=v^r$ while $H_d^{(0,q;r;s,t)}(u,v)=G_d^{(q;r;t)}(u,v)$.  Consequently,  $H_d^{(p,q;r;s,t)}(u,v)$ is expressible in terms of $G_d^{(q;r;t)}(u,v)$ as will be shown below.

It is possible to use the algebra of differential operators to derive the action of the derivatives on the functions $G_d^{(q;r;t)}(u,v)$ and $H_d^{(p,q;r;s,t)}(u,v)$.  One obtains

\vspace{-5pt}
\footnotesize
\eqna{
\D_{(u)}G_d^{(q;r;t)}(u,v)&=-\left[(r-t+q)\frac{u+v-1}{v}+t\right]G_d^{(q;r;t)}(u,v)+tG_d^{(q;r-1;t-1)}(u,v)\\
&\phantom{=}\hspace{0.5cm}-\frac{t(r-t+q)(r-t+q+1-d/2)}{(r-t)(r-t+1-d/2)}\frac{u}{v}G_d^{(q;r;t-1)}(u,v),\\
\D_{(v)}G_d^{(q;r;t)}(u,v)&=rG_d^{(q;r;t)}(u,v)-rG_d^{(q;r+1;t+1)}(u,v)+\frac{r(r-t+q)(r-t+q+1-d/2)}{(r-t)(r-t+1-d/2)}\frac{u}{v}G_d^{(q;r+1;t)}(u,v),
}
\normalsize
and

\vspace{-5pt}
\footnotesize
\eqna{
\D_{(u)}H_d^{(p,q;r;s,t)}(u,v)&=-\left[(r-s+p)\frac{u+v-1}{v}+s+q\right]H_d^{(p,q;r;s,t)}(u,v)+tH_d^{(p,q;r-1;s-1,t-1)}(u,v)\\
&\phantom{=}\hspace{0.5cm}-\frac{t(r-s+p)(r-t+q)(r-s+p+1-d/2)(r-t+q+1-d/2)}{(r-s)(r-t)(r-s+1-d/2)(r-t+1-d/2)}\frac{u}{v}H_d^{(p,q;r;s-1,t-1)}(u,v)\\
&\phantom{=}\hspace{1.0cm}+\frac{qt(q+s-t)(r-s+p)(r-s+p+1-d/2)}{(r-s)(r-t)(r-s+1-d/2)(r-t+1-d/2)}\frac{u}{v}H_d^{(p,q-1;r;s-1,t-1)}(u,v)\\
&\phantom{=}\hspace{1.5cm}-\frac{(q+s-t)(r-s+p)(r-s+p+1-d/2)}{(r-s)(r-s+1-d/2)}\frac{u}{v}H_d^{(p,q;r;s-1,t)}(u,v)\\
&\phantom{=}\hspace{2.0cm}+\frac{(q+s-t)(r-t-1)(r-t-d/2)}{(r-t+q-1)(r-t+q-d/2)}H_d^{(p,q;r-1;s-1,t)}(u,v)\\
&\phantom{=}\hspace{2.5cm}+\frac{q(q+s-t)(2r-t+q-1-d/2)}{(r-t+q-1)(r-t+q-d/2)}H_d^{(p,q-1;r;s,t)}(u,v)\\
&\phantom{=}\hspace{3.0cm}-\frac{qt(q+s-t)}{(r-t+q-1)(r-t+q-d/2)}H_d^{(p,q-1;r-1;s-1,t-1)}(u,v),\\
\D_{(v)}H_d^{(p,q;r;s,t)}(u,v)&=rH_d^{(p,q;r;s,t)}(u,v)-rH_d^{(p,q;r+1;s+1,t+1)}(u,v)\\
&\phantom{=}\hspace{0.5cm}+\frac{r(r-s+p)(r-t+q)(r-s+p+1-d/2)(r-t+q+1-d/2)}{(r-s)(r-t)(r-s+1-d/2)(r-t+1-d/2)}\frac{u}{v}H_d^{(p,q;r+1;s,t)}(u,v).
}
\normalsize
By using the knowledge of the derivatives and their actions on powers of $u$ and $u/v$, it is also possible to obtain several contiguous relations like
\vspace{-5pt}
\footnotesize
\eqna{
G_d^{(q+1;r;t)}(u,v)&=-\frac{t}{r-t}G_d^{(q;r;t-1)}(u,v)+\frac{r}{r-t}G_d^{(q;r+1;t)}(u,v)\\
&\phantom{=}\hspace{0.5cm}+\frac{rt(r-t+q+1)(r-t+q+2-d/2)}{(r-t)(r-t+1)(r-t+1-d/2)(r-t+2-d/2)}\frac{u}{v}G_d^{(q;r+1;t-1)}(u,v),\\
G_d^{(q;r-1;t)}(u,v)&=\frac{(r-t+q-1)(r-t+q-d/2)}{(r-t-1)(r-t-d/2)}\frac{1}{v}G_d^{(q;r;t)}(u,v)+\frac{qt}{(r-t-1)(r-t-d/2)}G_d^{(q-1;r-1;t-1)}(u,v)\\
&\phantom{=}\hspace{0.5cm}-\frac{q(2r-t+q-1-d/2)}{(r-t-1)(r-t-d/2)}G_d^{(q-1;r;t)}(u,v)\\
&\phantom{=}\hspace{1cm}-\frac{qt(r-t+q-1)(r-t+q-d/2)}{(r-t-1)(r-t)(r-t-d/2)(r-t+1-d/2)}\frac{u}{v}G_d^{(q-1;r;t-1)}(u,v),\\
G_d^{(q;r;t+1)}(u,v)&=\frac{(r-t+q-1)(r-t+q-d/2)}{(r-t-1)(r-t-d/2)}G_d^{(q;r;t)}(u,v)-\frac{q(r-2t+q-1-d/2)}{(r-t-1)(r-t-d/2)}G_d^{(q-1;r;t)}(u,v)\\
&\phantom{=}\hspace{0.5cm}-\frac{rq}{(r-t-1)(r-t-d/2)}G_d^{(q-1;r+1;t+1)}(u,v)\\
&\phantom{=}\hspace{1cm}+\frac{rq(r-t+q-1)(r-t+q-d/2)}{(r-t-1)(r-t)(r-t-d/2)(r-t+1-d/2)}\frac{u}{v}G_d^{(q-1;r+1;t)}(u,v).
}
\normalsize
Analogously, 
\vspace{-5pt}
\footnotesize
\eqna{
H_d^{(p+1,q;r;s,t)}(u,v)&=-\frac{(q+s-t)(r-t+q-s+1-d/2)}{(r-s)(r-s+1-d/2)}H_d^{(p,q;r;s-1,t)}(u,v)\\
&\phantom{=}\hspace{0.5cm}+\frac{r(q+s-t)(r-s+p+1)(r-t+q)(r-s+p+2-d/2)(r-t+q+1-d/2))}{(r-s)(r-s+1)(r-t)(r-s+1-d/2)(r-s+2-d/2)(r-t+1-d/2)}\\
&\phantom{=}\hspace{1.0cm}\times\frac{u}{v}H_d^{(p,q;r+1;s-1,t)}(u,v),\\
&\phantom{=}\hspace{1.5cm}-\frac{r(q+s-t)}{(r-s)(r-s+1-d/2)}H_d^{(p,q;r+1;s,t+1)}(u,v)\\
&\phantom{=}\hspace{2.0cm}+\frac{(r-t+q)(r-t+q+1-d/2)}{(r-s)(r-s+1-d/2)}H_d^{(p,q+1;r;s-1,t)}(u,v),\\
H_d^{(p,q+1;r;s,t)}(u,v)&=-\frac{t}{r-t}H_d^{(p,q;r;s,t-1)}(u,v)+\frac{r}{r-t}H_d^{(p,q;r+1;s+1,t)}(u,v)\\
&\phantom{=}\hspace{0.5cm}+\frac{rt(r-s+p)(r-t+q+1)(r-s+p+1-d/2)(r-t+q+2-d/2)}{(r-s)(r-t)(r-t+1)(r-s+1-d/2)(r-t+1-d/2)(r-t+2-d/2)}\\
&\phantom{=}\hspace{1cm}\times\frac{u}{v}H_d^{(p,q;r+1;s,t-1)}(u,v),\\
H_d^{(p,q;r;s+1,t)}(u,v)&=\frac{(r-s+p-1)(r-s+p-d/2)}{(r-s-1)(r-s-d/2)}H_d^{(p,q;r;s,t)}(u,v)-\frac{p(r-2s+p-1-d/2)}{(r-s-1)(r-s-d/2)}H_d^{(p-1,q;r;s,t)}(u,v)\\
&\phantom{=}\hspace{0.5cm}+\frac{pr(r-s+p-1)(r-t+q)(r-s+p-d/2)(r-t+q+1-d/2)}{(r-s)(r-s-1)(r-t)(r-s+1-d/2)(r-s-d/2)(r-t+1-d/2)}\\
&\phantom{=}\hspace{1.0cm}\times\frac{u}{v}H_d^{(p-1,q;r+1;s,t)}(u,v)-\frac{pr}{(r-s-1)(r-s-d/2)}H_d^{(p-1,q;r+1;s+1,t+1)}(u,v),\\
H_d^{(p,q;r;s,t+1)}(u,v)&=\frac{(r-t+q-1)(r-t+q-d/2)}{(r-t-1)(r-t-d/2)}H_d^{(p,q;r;s,t)}(u,v)-\frac{q(r-2t+q-1-d/2)}{(r-t-1)(r-t-d/2)}H_d^{(p,q-1;r;s,t)}(u,v)\\
&\phantom{=}\hspace{0.5cm}+\frac{qr(r-s+p)(r-t+q-1)(r-s+p+1-d/2)(r-t+q-d/2)}{(r-s)(r-t)(r-t-1)(r-s+1-d/2)(r-t+1-d/2)(r-t-d/2)}\\
&\phantom{=}\hspace{1.0cm}\times\frac{u}{v}H_d^{(p,q-1;r+1;s,t)}(u,v)-\frac{qr}{(r-t-1)(r-t-d/2)}H_d^{(p,q-1;r+1;s+1,t+1)}(u,v),
}
\normalsize
where the contiguous relation for $H_d^{(p,q;r-1;s,t)}(u,v)$ is omitted since it is quite complicated.

%%%%%%%%%%%%%%%%%%%%%%%%%%%%%%%%%%%%%%%%%%%%%%%%%%
%%%%%%%%%%%%%%%%%%%%%%%%%%%%%%%%%%%%%%%%%%%%%%%%%%

\subsection{Power Series}

Functional form of  $G_d^{(q;r;t)}(u,v)$ and $H_d^{(p,q;r;s,t)}(u,v)$ can be obtained by expressing them as multiple sums over powers of the variables $x=u/v$ and $y=1-1/v$.  First, using the recurrence relation \eqref{EqRR} with $k=1$, it is easy to verify that
\eqn{G_d^{(q;r;t)}(u,v)=\sum_{m,n\geq0}\frac{(-q)_m(-t)_m}{(r-t+1-d/2)_mm!}\frac{(r)_{m+n}(r-t+q)_{m+n}}{(r-t)_{2m+n}n!}x^my^n.}[EqG]
In terms of the hypergeometric function $G(\alpha,\beta,\gamma,\delta;x,y)$ of Exton~\cite{0305-4470-28-3-017,Dolan:2003hv}, \eqref{EqG} can be expressed as
\eqn{G_d^{(q;r;t)}(u,v)=G(r,r-t+q,r-t+1-d/2,r-t;x,y).}

To obtain $H_d^{(p,q;r;s,t)}(u,v)$, we start from \eqref{EqHinG} using $G(\alpha,\beta,\gamma,\delta;x,y)$ in \eqref{EqG} rewritten in terms of the fourth Appell hypergeometric function. Then we apply the derivatives $\mathcal{D}_{(u,v)}$ as done for $G_d^{(q;r;t)}(u,v)$.  The result is given in terms of four infinite sums: two sums of functions $G_d^{(q;r;t)}(u,v)$ that itself is given in terms of a double sum.  This expression can be simplified, with the help of generalizations of Gauss' identity, to an expression with two infinite sums and two finite sums given by

\vspace{-5pt}
\footnotesize
\eqna{
H_d^{(p,q;r;s,t)}(u,v)&=\sum_{m,n\geq0}P_d^{(p,q;r;s,t)}(m,n)\frac{(r)_{m+n}(r-s+p)_{m+n}(r-t+q)_{m+n}}{(r-s)_{2m+n}(r-s+1-d/2)_m(r-t)_{2m+n}(r-t+1-d/2)_mn!}x^my^n,\\
P_d^{(p,q;r;s,t)}(m,n)&=\sum_{i=0}^m\sum_{j=0}^i\frac{(-i)_j}{i!j!(m-i)!}(-p)_i(-q)_{m-i+j}(-s-q+m-i+j)_{i-j}(-t)_{m-i+j}(r-s+m+n+i)_{m-i}\\
&\phantom{=}\hspace{0.5cm}\times(r-s+p+1-d/2)_{m-i}(r-t+2m+n-i+j)_{i-j}(r-t+1-d/2+m-i)_i.
}[EqH]
\normalsize
$P_d^{(p,q;r;s,t)}(m,n)$ is a polynomial of degree $m$  in the variables $\{p,q,d\}$ when each of these variables is considered separately and a polynomial of degree $2m$ in the variables $\{r,s,t\}$, again considering  each of these variables independently.  Moreover, $P_d^{(p,q;r;s,t)}(m,n)$  is written as a sum over $(m+1)(m+2)/2$ terms, however both sums in $P_d^{(p,q;r;s,t)}(m,n)$ can be extended to infinity as the expression vanishes for $i>m$ or $j>i$.  From the recurrence relation \eqref{EqRR} with $k=1$, it satisfies

\vspace{-5pt}
\footnotesize
\eqna{
P_d^{(p+1,q;r;s,t)}(m,n)&=\frac{r-s+p+1-d/2+m}{r-s+p+1-d/2}P_d^{(p,q;r;s,t)}(m,n)\\
&\phantom{=}\hspace{0.5cm}-\frac{(r-s-1+2m+n)(r-s-d/2+m)(r-t-1+2m+n)(r-t-d/2+m)}{r-s+p+1-d/2}\\
&\phantom{=}\hspace{1.0cm}\times P_d^{(p,q;r;s,t)}(m-1,n+1)\\
&\phantom{=}\hspace{1.5cm}+\frac{(r+m+n)(r-s-d/2+m)(r-t-d/2+m)(r-t+q+m+n)}{r-s+p+1-d/2}\\
&\phantom{=}\hspace{2.0cm}\times P_d^{(p,q;r;s,t)}(m-1,n+2),
}
\normalsize
by construction.

It is important to note that \eqref{EqG} and \eqref{EqH} fulfils all the expected limits.  Indeed, $G_d^{(q;r;t)}(0,1)=1$ and $G_d^{(0;r;t)}(u,v)=v^r$, while $H_d^{(p,q;r;s,t)}(0,1)=1$ and $H_d^{(0,q;r;s,t)}(u,v)=G_d^{(q;r;t)}(u,v)$.  Moreover, by examining \eqref{EqG} it is straightforward to check the $G_d^{(t;r-t+q;q)}(u,v)=G_d^{(q;r;t)}(u,v)$ symmetry relation.  Hence, by construction, \eqref{EqH} satisfies all the symmetry properties.  The symmetry $H_d^{(p,t;r-t+q;s-t+q,q)}(u,v)=H_d^{(p,q;r;s,t)}(u,v)$ is also straightforward to verify, but the symmetry $H_d^{(q,p;r;t,s)}(u,v)=H_d^{(p,q;r;s,t)}(u,v)$ is not.

%%%%%%%%%%%%%%%%%%%%%%%%%%%%%%%%%%%%%%%%%%%%%%%%%%
%%%%%%%%%%%%%%%%%%%%%%%%%%%%%%%%%%%%%%%%%%%%%%%%%%

\subsection{Discrete Invariance}

We now point out that $H_d^{(p,q;r;s,t)}(u,v)$, and consequently $P_d^{(p,q;r;s,t)}(m,n)$, are invariant under the dihedral group $D_6$ of order $12$,
\eqna{
H_d^{(p,q;r;s,t)}(u,v)&=H_d^{g\cdot(p,q;r;s,t)}(u,v),\\
P_d^{(p,q;r;s,t)}(m,n)&=P_d^{g\cdot(p,q;r;s,t)}(m,n),  
}
where $g\in D_6$, since the two symmetries discussed above generate the dihedral group $D_6$. 

Indeed, defining the action of the group $g\cdot(p,q;r;s,t)$ as standard matrix multiplication on the five-dimensional vector $(p,q,r,s,t)^T$, the two symmetries are represented by the following matrices
\begin{gather*}
t_1=\left(\begin{array}{ccccc}0&1&0&0&0\\1&0&0&0&0\\0&0&1&0&0\\0&0&0&0&1\\0&0&0&1&0\end{array}\right),\quad\quad t_2=\left(\begin{array}{ccccc}1&0&0&0&0\\0&0&0&0&1\\0&1&1&0&-1\\0&1&0&1&-1\\0&1&0&0&0\end{array}\right).
\end{gather*}

Hence, $x=t_1$ and $y=t_2t_1$ lead to the presentation $\langle x,y|x^2=y^6=(xy)^2=1\rangle$ of the dihedral group $D_6$.  By defining rotations $r_i=y^i$ and reflections $s_i=xy^{-i}$, it is easy to observe that
\eqn{r_ir_j=r_{i+j},\quad\quad r_is_j=s_{i+j},\quad\quad s_ir_j=s_{i-j},\quad\quad s_is_j=r_{i-j},}
which are the correct multiplication rules of $D_6$.  Since $D_6$ has six conjugacy classes given by
\begin{gather*}
E=\{r_0\},\quad\quad C_6=\{r_1,r_5\},\quad\quad C_3=\{r_2,r_4\},\\
C_2=\{r_3\},\quad\quad C_2'=\{s_0,s_2,s_4\},\quad\quad C_2''=\{s_1,s_3,s_5\},
\end{gather*}
it has six irreducible representations ($A_{1,2}$, $B_{1,2}$ and $E_{1,2}$) of dimensions given by the character of the conjugacy class $E$ shown in table \ref{TabChar}.
\begin{table}[t]
\centering
\begin{tabular}{|c|c|c|c|c|c|c|}
\hline
 & $E$ & $C_6$ & $C_3$ & $C_2$ & $C_2'$ & $C_2''$\\\hline
$A_1$ & $1$ & $1$ & $1$ & $1$ & $1$ & $1$\\\hline
$A_2$ & $1$ & $1$ & $1$ & $1$ & $-1$ & $-1$\\\hline
$B_1$ & $1$ & $-1$ & $1$ & $-1$ & $1$ & $-1$\\\hline
$B_2$ & $1$ & $-1$ & $1$ & $-1$ & $-1$ & $1$\\\hline
$E_1$ & $2$ & $1$ & $-1$ & $-2$ & $0$ & $0$\\\hline
$E_2$ & $2$ & $-1$ & $-1$ & $2$ & $0$ & $0$\\
\hline
\end{tabular}
\caption{Table of characters for dihedral group $D_6$.}
\label{TabChar}
\end{table}
The five-dimensional reducible representation over the space of $(p,q;r;s,t)$ decomposes as two one-dimensional irreducible representation $A_1$, one one-dimensional irreducible representation $B_2$, and one two-dimensional irreducible representation $E_2$.  A possible similarity transformation matrix $S$ is
\eqn{S=\left(\begin{array}{ccccc}1&1&1&0&0\\0&0&2&-1&-1\\0&0&0&1&-1\\1&0&0&-1&0\\0&1&0&0&-1\end{array}\right),}
which makes $Sr_iS^{-1}$ and $Ss_iS^{-1}$ block-diagonal for all $i$.  Therefore, under group transformations, the linear combinations $p+q+r$ and $2r-s-t$ are invariant, while the linear combination $s-t$ changes sign under reflections.  The remaining linear combinations, $p-s$ and $q-t$, transform according to the action of $E_2$.  From the invariant theory of finite groups, the polynomial $P_d^{(p,q;r;s,t)}(m,n)$ and the function $H_d^{(p,q;r;s,t)}(u,v)$ are thus functions of $p+q+r$, $2r-s-t$, $(s-t)^2$, $(p-s)^2-(p-s)(q-t)+(q-t)^2$ and $[(p-s)+(q-t)][2(p-s)-(q-t)][(p-s)-2(q-t)]$ although it is not straightforward to verify from \eqref{EqH}. As will be shown in a forthcoming publication, some of the well-behaved linear combinations found here have direct links to the conformal blocks while others are useful to simplify results.

%%%%%%%%%%%%%%%%%%%%%%%%%%%%%%%%%%%%%%%%%%%%%%%%%%
%%%%%%%%%%%%%%%%%%%%%%%%%%%%%%%%%%%%%%%%%%%%%%%%%%

\subsection{Relation to Conformal Blocks}

As mentioned before, conformal blocks can be obtained from linear combinations of the function $H_d^{(p,q;r;s,t)}(u,v)$ with extra vector derivatives (fractional calculus is not needed anymore).  The main purpose in computing the function $H_d^{(p,q;r;s,t)}(u,v)$ is that it has the proper analytic behavior to generate conformal blocks.  Indeed, all scalar derivatives with real powers have been evaluated correctly.

To connect with the extensive literature on conformal blocks (see \textit{e.g.} \cite{Costa:2011mg,Costa:2011dw,Kos:2014bka,Costa:2014rya,Rejon-Barrera:2015bpa,Penedones:2015aga,Echeverri:2016dun,Costa:2016hju,Schomerus:2016epl}), we give here two linear combinations corresponding to scalar and spin one exchange in the four-point correlation function of four scalars.  Writing the conformal blocks as $G_{\Delta,J}^{\Delta_1,\Delta_2,\Delta_3,\Delta_4}(u,v)$ for four scalar quasi-primary operators with conformal dimensions $\Delta_{1,\ldots,4}$ and an exchanged quasi-primary operator with conformal dimension $\Delta$ and spin $J$, they are given by

\vspace{-5pt}
\footnotesize
\eqna{
G_{\Delta,J=0}^{\Delta_1,\Delta_2,\Delta_3,\Delta_4}(u,v)&=v^{(\Delta_{ij}-\Delta)/2}H_d^{\left(-\frac{\Delta_{12}+\Delta}{2},-\frac{\Delta_{34}+\Delta}{2};\Delta;0,0\right)}(u,v),\\
G_{\Delta,J=1}^{\Delta_1,\Delta_2,\Delta_3,\Delta_4}(u,v)&=\frac{4\Delta^2v^{(\Delta_{ij}-\Delta+1)/2}}{(\Delta_{12}-\Delta+1)(\Delta_{34}-\Delta+1)}\\
&\phantom{=}\hspace{20pt}\times\left[H_d^{\left(-\frac{\Delta_{12}+\Delta+1}{2},-\frac{\Delta_{34}+\Delta+1}{2};\Delta+1;1,1\right)}(u,v)-H_d^{\left(-\frac{\Delta_{12}+\Delta+1}{2},-\frac{\Delta_{34}+\Delta+1}{2};\Delta;0,0\right)}(u,v)\right]\\
&\phantom{=}\hspace{40pt}+\frac{(\Delta+1)(\Delta+1-d-\Delta_{12})(\Delta+1-d-\Delta_{34})uv^{(\Delta_{ij}-\Delta-1)/2}}{4(\Delta+1-d/2)^2(\Delta+1-d)}\\
&\phantom{=}\hspace{60pt}\times H_d^{\left(-\frac{\Delta_{12}+\Delta+1}{2},-\frac{\Delta_{34}+\Delta+1}{2};\Delta+1;0,0\right)}(u,v),
}
\normalsize
where $\Delta_{ij}=\Delta_i-\Delta_j$.  They correspond to the conformal blocks of Dolan and Osborn \cite{Dolan:2000ut,Dolan:2003hv}.  To obtain these results, the general strategy is to consider the three-point correlation functions and contract the indices appropriately.  By doing so, the remaining derivatives are all scalar, and they act on linear combinations of products of embedding space coordinates, like in \eqref{EqFpq}.  It is then straightforward to normalize the results to match known conformal blocks.  It is important to point out here that extra derivatives were not needed to generate the vector exchange conformal block.  However, that is not the case for most conformal blocks as generalization to any spin is not straightforward.  The study of these more complicated blocks will be discussed elsewhere.

%%%%%%%%%%%%%%%%%%%%%%%%%%%%%%%%%%%%%%%%%%%%%%%%%%
%%%%%%%%%%%%%%%%%%%%%%%%%%%%%%%%%%%%%%%%%%%%%%%%%%

\section{Summary}\label{SecSum}

We showed several technical results that are needed to implement the OPE and to calculate conformal blocks in embedding space. We anticipate that these results are a necessary step to developing a new method for calculating conformal blocks, but several further advances are still required.  The right-hand side of the OPE contains sums over quasi-primary operators and their descendants. We were able to reduce all non-trivially acting differential operators to a single one that can appear in the OPE in order to generate descendants. When the basic differential operator is used in the four-point function it is more natural to express all results in terms of the conformal invariants, $u$ and $v$, therefore we presented relevant expressions in term of those variables as well. 

We demonstrated how the differential operator can be used to obtain a function in terms of which conformal blocks can be constructed. This follows from the action of derivative operators in two OPEs that one uses to reduce a four-point function. The most compact expression that we were able to obtain for the function needed to build conformal blocks is written as a quadruple sum in \eqref{EqH}. Two of these sums are infinite and they are over powers of two combinations of the conformal ratios $u/v$ and $1-1/v$. The remaining two sums are finite and involve only numerical constants that depend on the quantum numbers of operators. There are many interesting properties that the conformal block master function satisfies, among those an invariance under the dihedral group $D_6$. Our expressions do not yet contain un-contracted Lorentz indices corresponding to Lorentz quantum numbers of external fields, but these can be obtained with more derivative operators. Thus, conformal blocks will be expressed in terms of the master function and its derivatives. 

A future publication will address the details of obtaining the OPE in our formalism and of constructing conformal blocks for arbitrary Lorentz quantum numbers. An interesting followup would be investigating supersymmetric version of this program given the formalism in \cite{Goldberger:2011yp,Goldberger:2012xb}. Interestingly, \cite{Isachenkov:2016gim} points to a supersymmetric version of the Casimir equation and their solutions which may be very useful for superconformal field theories.  Finally, from the correspondence found in \cite{Isachenkov:2016gim} between conformal blocks and integrable systems, our master function may be of interest in the analysis of Calogero-Sutherland Hamiltonians.

%%%%%%%%%%%%%%%%%%%%%%%%%%%%%%%%%%%%%%%%%%%%%%%%%%
%%%%%%%%%%%%%%%%%%%%%%%%%%%%%%%%%%%%%%%%%%%%%%%%%%

\ack{
The authors would like to thank Pierre Mathieu for useful discussions and the CERN Theory Group, where this work was conceived, for its hospitality.  The work of JFF is supported by NSERC.  WS is supported in part by the U.~S.~Department of Energy under the contract DE-FG02-92ER-40704.
}

%%%%%%%%%%%%%%%%%%%%%%%%%%%%%%%%%%%%%%%%%%%%%%%%%%
%%%%%%%%%%%%%%%%%%%%%%%%%%%%%%%%%%%%%%%%%%%%%%%%%%

\bibliography{DiffOp.bbl}

\end{document}